\documentclass[twocolumn,times]{aastex631}

\usepackage{amsmath}	        
\usepackage{textcomp,gensymb}	
\usepackage{xcolor}             
\usepackage{mathtools}

\newcommand{\thisstar}{HD~12800}

\newcommand\ms{\ensuremath{\text{m}\,\text{s}^{-1}}}

\definecolor{my_color}{HTML}{CF0000}
\newcommand\bmaroon{}

\received{18 July, 2025}
\revised{16 February, 2026}
\accepted{4 March, 2026}

\begin{document}

\title{A Planetary Illusion's Funeral:
Non-detection of a \textit{Gaia} DR3 Exoplanet Candidate, and the Role of Intermediate-precision Radial Velocities in \textit{Gaia} Exoplanet Follow-up\footnote{Read as ``A Planetary Fantasy"}
}

\correspondingauthor{Alexander Venner}
\email{alvenner@mpia.de}

\author[0000-0002-8400-1646]{Alexander Venner}
\affiliation{Centre for Astrophysics, University of Southern Queensland, Toowoomba, QLD 4350, Australia}
\affiliation{Max Planck Institute for Astronomy, 69117 Heidelberg, Germany}

\author[0000-0003-0918-7484]{Chelsea X. Huang}
\affiliation{Centre for Astrophysics, University of Southern Queensland, Toowoomba, QLD 4350, Australia}

\author[0000-0001-9911-7388]{David W. Latham}
\affiliation{Center for Astrophysics, Harvard \& Smithsonian, Cambridge, MA 02138, USA}

\author[0000-0002-8964-8377]{Samuel N. Quinn}
\affiliation{Center for Astrophysics, Harvard \& Smithsonian, Cambridge, MA 02138, USA}

\author[0000-0001-6637-5401]{Allyson Bieryla}
\affiliation{Center for Astrophysics, Harvard \& Smithsonian, Cambridge, MA 02138, USA}
\affiliation{Centre for Astrophysics, University of Southern Queensland, Toowoomba, QLD 4350, Australia}

\author[0000-0001-7246-5438]{Andrew Vanderburg}
\affiliation{Center for Astrophysics, Harvard \& Smithsonian, Cambridge, MA 02138, USA}

\author[0000-0001-9957-9304]{Robert A. Wittenmyer}
\affiliation{Centre for Astrophysics, University of Southern Queensland, Toowoomba, QLD 4350, Australia}



\begin{abstract}

The detection of exoplanets using astrometry has long been an area of interest, but is fraught with challenges. The \textit{Gaia} mission is fundamentally reshaping this field thanks to its unprecedentedly precise all-sky astrometric observations. The 2022 release of \textit{Gaia}~DR3 brought the first exoplanets discovered from the \textit{Gaia} astrometry, including a new candidate around the bright ($V=6.6$) solar-type star HD~12800. However, two years after announcement, the \textit{Gaia} exoplanet candidate was retracted. In this work we report radial velocity observations of HD 12800 acquired with the TRES spectrograph, which we began immediately after the release of \textit{Gaia}~DR3. Our observations failed to detect the planet candidate; nonetheless, we emphasise that the originally proposed companion would have been easily detected in our radial velocity observations. We conclude with a discussion on the role of intermediate-precision ($\approx$10~\ms{}) RV spectrographs in the follow-up of \textit{Gaia} astrometric exoplanet candidates, relevant to the forthcoming release of \textit{Gaia}~DR4. We argue that such observations may play an important role in planet confirmation for stars between approximately $8<G<12$, likely to represent a significant fraction of \textit{Gaia} exoplanet discoveries.


\end{abstract}



\section{Introduction} \label{sec:intro}

Many of the methods used for the discovery of exoplanets involve the detection of the subtle influence they have on their parent stars. Perhaps the \bmaroon{earliest} of these methods \bmaroon{to be applied} is astrometry, which may be used to indirectly detect the existence of planets through the detection of the minute reflex orbital motion of their host stars as they move across the sky \citep{Sozzetti2005, Quirrenbach2010}.

Though the discovery of exoplanets through astrometry has been explored for a long time, the method has unfortunately enjoyed rather few successes. Ground-based detections have been attempted since the middle of the twentieth century \bmaroon{(e.g. \citealt{Lippincott1960, vandeKamp1963, vandeKamp1969, Hershey1973, Gatewood1973, Gatewood1974}}, but the level of precision required means that for most stars planet detection is only possible with space-based observations \bmaroon{(see \citealt{Sozzetti2005}, section~4.1 for a review)}. Even in the case of the \textit{Hipparcos} mission \citep{Hipparcos}, only a small number of planets then known could be detected \citep{Reffert2011}. 

However, this state of affairs is changing thanks to the recently-completed \textit{Gaia} mission \citep{Gaia}. \textit{Gaia} has performed a deep all-sky astrometric survey with unprecedented precision and duration. Early-mission yield estimates set the expectation that \textit{Gaia} will astrometrically detect in excess of $>$10$^4$ exoplanets \citep{Perryman2014}, a considerably larger count than the entire sample of exoplanets currently known. There is therefore significant anticipation for the exoplanet science results from the 5-year \textit{Gaia} nominal mission, included in \textit{Gaia} Data Release~4 (DR4) which is expected in December~2026.\footnote{\url{https://www.cosmos.esa.int/web/gaia/release}}

\textit{Gaia}~DR3, released in June~2022 based on the first three years of \textit{Gaia} observations \citep{GaiaDR3}, included the first batch of non-single star (NSS) orbital solutions from the mission. The detection pipeline and the properties of the orbital solutions are primarily discussed in \citet{Arenou2023}.
The \textit{Gaia}~DR3 NSS solutions have already proved fruitful for exoplanet research. \citet{Winn2022} explored joint constraints on exoplanet parameters by combining the \textit{Gaia} astrometric orbital solutions with archival RV observations, demonstrating consistency between solutions in the best cases. Follow-up RV observations of the star HIP~66074 led to the first confirmation of an exoplanet \bmaroon{reported} in the \textit{Gaia} astrometry, Gaia-3~b \citep{Sozzetti2023}. Most recently, Gaia-4~b and Gaia-5~b have been confirmed as massive companions (10 -- 20$~M_J$) orbiting low-mass stars \citep{Stefansson2025}\bmaroon{, while Gaia-6~B is a confirmed long-period and high-eccentricity companion, though with significantly revised parameters \citep{Pinamonti2026}.}

However, much care must be taken to identify false positives among the \textit{Gaia} astrometric exoplanet candidates. The main astrophysical false positive scenario for these systems involve binary systems with near-equal luminosities. As \textit{Gaia} observes the motion of the centre-of-light much smaller than the angular resolution of the instrument, for a binary where both components are luminous in the \textit{Gaia} $G$-band the orbital motion will be attenuated according to the flux ratio.
The first system to be recognised as such a false positive is HD~68683, identified in \citet{Holl2023} on the basis of published spectroscopic observations that show it is a double-lined spectroscopic binary (SB2). Three additional \textit{Gaia}~DR3 candidates were discovered to be SB2s in \citet{Marcussen2023}, demonstrating that these false positives are not uncommon but can easily be detected with spectroscopic follow-up. It is also possible to detect these false positives through imaging; HD~3221, highlighted as a planet candidate orbiting a young star in \citet{Arenou2023}, was independently resolved as a near-equal luminosity binary by \citet{Bonavita2022}.

A more pathological false positive scenario involves fully spurious orbital solutions. At the time of writing, a total of four \textit{Gaia}~DR3 NSS orbits have been identified as spurious and have been retracted, being caused by software issues.\footnote{\url{https://www.cosmos.esa.int/web/gaia/dr3-known-issues\#FalsePositive}} These manifested as systems with NSS astrometric solutions that failed to demonstrate the expected signals of either planetary or stellar companions in follow-up observations. The first of these to be reported was HD~113283, with a stellar companion reported in \textit{Gaia}~DR3 that went undetected in follow-up RV observations \citep{Spaeth2023}. Another casualty was the candidate companion of WD~0141-675, the only planet candidate orbiting a white dwarf detected in \textit{Gaia}~DR3 (\citealt{Arenou2023}, section~8.8; see further \citealt{Rogers2024}). \bmaroon{The NSS solution for HIP~66074 (Gaia-3) was also retracted, which appears unexpected given the earlier confirmation of a planet with a similar orbital period in \citet{Sozzetti2023} and complicates its claim to being the first exoplanet to be discovered from \textit{Gaia} astrometry.}

\bmaroon{The last} of the \bmaroon{four} now-retracted exoplanet candidates pertains to the bright ($V=6.6$) solar-type star \thisstar{}. The NSS solution was highlighted as ``the only candidate companion around a main-sequence solar-type star with a mass well in the planetary regime" newly reported from the \textit{Gaia}~DR3 targeted search \citep[][section~8.7]{Arenou2023}, with orbital period $P=401\pm12$~days, orbital semi-major axis $a_0=0.25\pm0.05$~milli-arcseconds (mas), and estimated mass $5.6\pm1.4~M_J$. However, it was ultimately discovered that the two-body solution was spurious, and the \textit{Gaia}~DR3 NSS solution was retracted on 2024-05-27.

Soon after the release of \textit{Gaia}~DR3, we identified the candidate companion of \thisstar{} as an interesting target for follow-up and immediately began collecting RV data. However, our observations failed to detect the expected orbital signal, which demonstrated that the candidate companion did not exist prior to its formal retraction. In this work, we report on our observations of \thisstar{} and their implications for follow-up of \textit{Gaia} astrometric exoplanet candidates in the context of the future release of \textit{Gaia}~DR4.

\section{Target Information} \label{sec:target}

\begin{deluxetable*}{lrr}[ht!]
\label{tab:parameters}
\centering
\tablecaption{Properties of \thisstar{}.}
\tablehead{\colhead{Parameter} & \colhead{Value} & \colhead{Reference}}
\startdata
~~Right Ascension $\alpha_{\text{J2000}}$ \dotfill & 02:09:08.26 & \citet{GaiaDR3} \\
~~Declination $\delta_{\text{J2000}}$ \dotfill & +71:33:07.22 & \citet{GaiaDR3} \\
~~Parallax $\varpi$ (mas) \dotfill & $37.012\pm0.017$ & \citet{GaiaDR3} \\
~~Distance $d$ (pc) \dotfill & $27.018\pm0.012$ & \citet{GaiaDR3} \\
~~$V$ (mag) \dotfill & $6.58\pm0.01$ & \citet{Tycho2} \\
\hline
~~$B_\text{T}$ (mag) \dotfill & $7.225\pm0.015$ & \citet{Tycho2} \\
~~$V_\text{T}$ (mag) \dotfill & $6.633\pm0.010$ & \citet{Tycho2} \\
~~$G$ (mag) \dotfill & $6.448\pm0.02$ & \citet{GaiaDR3} \\
~~$G_{\text{BP}}$ (mag) \dotfill & $6.717\pm0.02$ & \citet{GaiaDR3} \\
~~$G_{\text{RP}}$ (mag) \dotfill & $6.009\pm0.02$ & \citet{GaiaDR3} \\
~~$J$ (mag) \dotfill & $5.519\pm0.027$ & \citet{2MASS} \\
~~$H$ (mag) \dotfill & $5.313\pm0.038$ & \citet{2MASS} \\
~~$K_S$ (mag) \dotfill & $5.241\pm0.017$ & \citet{2MASS} \\
\hline
~~$T_\text{eff}$ (K) \dotfill & $6000\pm50$ & This work \\
~~$\log g$ ($\log$~c\ms{}) \dotfill & $4.26\pm0.10$ & This work \\
~~$[\text{M/H}]$ (dex) \dotfill & $-0.24\pm0.08$ & This work \\
~~$v_\text{broad}$ (k\ms{}) * \dotfill & $4.0\pm0.5$ & This work \\
~~$M_*$ ($M_\odot$) \dotfill & $0.98^{+0.05}_{-0.04}$ & This work \\
~~$R_*$ ($R_\odot$) \dotfill & $1.082\pm0.024$ & This work \\
~~$L_*$ ($L_\odot$) \dotfill & $1.36\pm0.07$ & This work \\
~~Age (Gyr) \dotfill & $5.7^{+1.7}_{-1.9}$ & This work \\
\enddata
\tablecomments{(*) Includes both $v\sin i$ and macroturbulence.}
\end{deluxetable*}
\vspace{-9 mm}

\thisstar{} (54~Cas, Gaia~DR3~522135261462534528) is a $V=6.6$ star at a distance of $27.018\pm0.012$~parsecs \citep{GaiaDR3} with a spectral type conventionally given as F8\bmaroon{, originating from the Henry Draper catalogue \citep[see][]{Cannon1918}}. Despite its status as a bright and nearby Sun-like star, \thisstar{} has little presence in the astronomical literature and has largely avoided study.

In \textit{Gaia}~DR3 the star had a \texttt{OrbitalTargetedSearch} solution in the \texttt{nss\_two\_body\_orbit} table. Though the two-body solution has now been retracted, we go through the process of extracting the orbital information from the \textit{Gaia} orbital solution in order to make full comparisons with our RV observations. The \textit{Gaia}~DR3 solution has an orbital period of $P=401\pm12$~days, eccentricity $e=0.22\pm0.16$, time of periastron $T_\text{p}=118\pm69$~days (i.e. BJD $2457507\pm69$ relative to epoch $2016.0$), and the following Thiele-Innes orbital coefficients: $A=0.157\pm0.096$, $B=0.020\pm0.079$, $F=-0.152\pm0.087$, and $G=-0.140\pm0.051$. To convert the reported Thiele-Innes coefficients to the corresponding Campbell terms, we used the \texttt{nsstools} package \citep[see][appendices~A,~B]{Halbwachs2023}.\footnote{\url{https://www.cosmos.esa.int/web/gaia/dr3-nss-tools}} This provides a semi-major axis of $a_0=0.25\pm0.05$~mas, argument of periastron $\omega=54\pm32\degree$, orbital inclination $i=107.8\pm8.3\degree$, and longitude of node $\Omega=30.4\pm5.7\degree$ for the \textit{Gaia}~DR3 two-body solution. These values align with those reproduced in \citet{Arenou2023}.

In the case of a two-body orbit where the secondary contributes no flux, as appropriate for a star-planet system, the observable amplitude of the astrometric orbit $a_0$ is related to the system properties through the following equation:

\begin{equation}
    \label{eq:astrometry}
    a_0=\frac{M_\text{p}}{M_*}\frac{a}{D}
    \;,
\end{equation}

where $a_0$ is in milli-arcseconds, $M_*$ and $M_\text{p}$ are the mass of the star and companion respectively, $a$ is the orbital semi-major axis in AU, and $D$ is the distance in parsecs \citep[][equation~6]{Sozzetti2005}. Assuming for the moment $M_*=1~M_\odot$, the $a_0=0.25\pm0.05$~mas originally reported in \textit{Gaia}~DR3 entails a companion mass of $M_\text{p}\approx6.3~M_J$.

We may then estimate the expected radial velocity semi-amplitude $K$ using the classical expression:

\begin{equation}
    \label{eq:RV}
    K=\sqrt{\frac{G}{(M_*+M_\text{p})a(1-e^2)}}M_\text{p}\sin(i)
\end{equation}

Where $G$ is the gravitational constant \citep[][equation~12]{Lovis2010}. We provisionally estimated a reflex semi-amplitude of $K\approx170$~\ms{}, which notionally could be easily detected in RV observations. This motivated our attempt to detect the RV signal from the planet candidate.

\section{Observations and Results}  \label{sec:observations}

\begin{figure*}[t!]
    \centering
    \includegraphics[width=\textwidth]{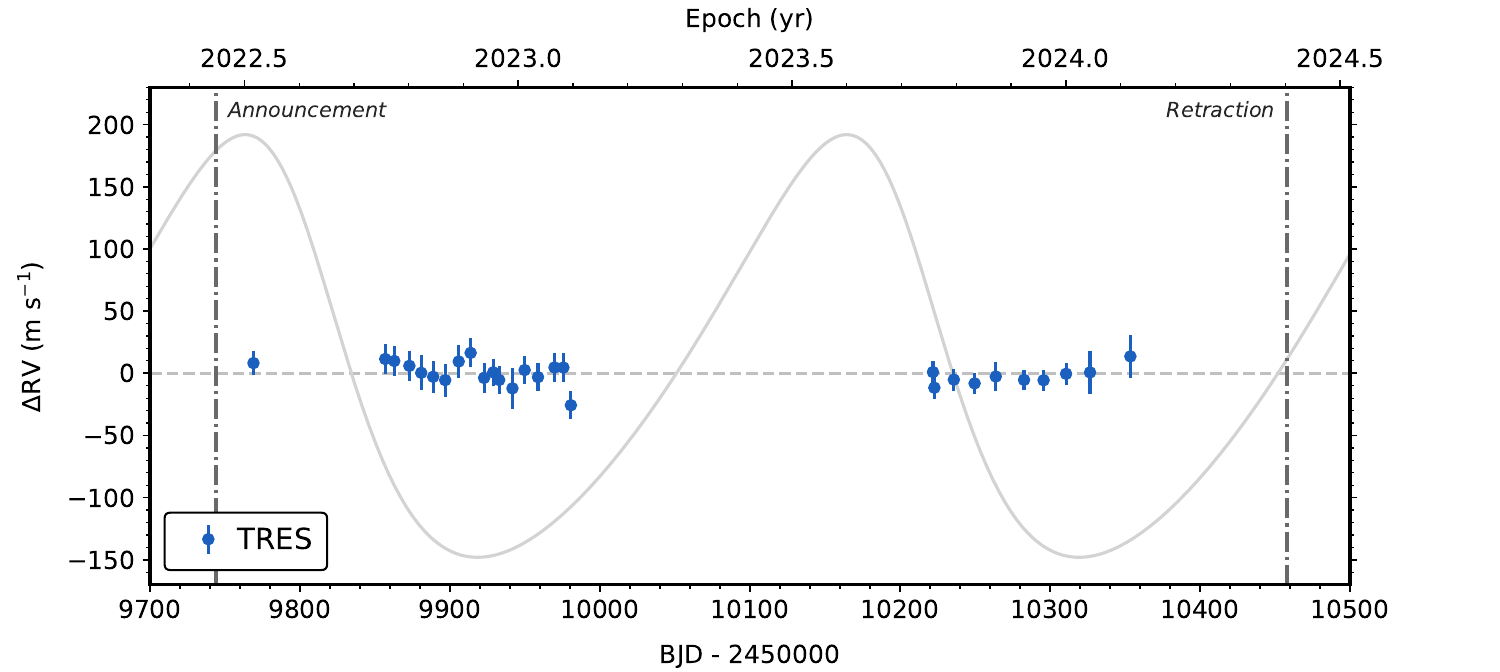}
    \caption{TRES RV observations of \thisstar{}. We mark the dates of announcement (2022-06-13) and official retraction (2024-05-27) for the \textit{Gaia}~DR3 planet candidate. Our observations would have easily detected the RV signal expected from the \textit{Gaia}~DR3 orbital solution, and independently \bmaroon{verify} that the companion does not exist.
    While in this case the ultimate result is a non-detection, this nonetheless serves to demonstrate the value of intermediate-precision ($\approx$10~\ms{}) RV observations for the follow-up of \textit{Gaia} astrometric exoplanet candidates.}
    \label{fig:RV}
\end{figure*}

Though \thisstar{} is a bright Sun-like star, it happens to have been omitted from the major long-term exoplanet RV surveys, perhaps on account of its relatively blue colours \citep[$B-V=0.59$,][]{Tycho2}. The most recent RV data published in the literature are low-precision observations ($\sigma=0.33$~k\ms{}) from \citet{Latham2002}, though the star has occasionally been used as a spectral standard in subsequent literature \citep[ex.][]{Tingley2011}. We report the relevant observational parameters in Table~\ref{tab:parameters}.

We therefore began collecting new observations of \thisstar{} soon after the 2022-06-13 release of \textit{Gaia}~DR3. We observed the star using the TRES spectrograph on the 1.5~m telescope at the Fred Lawrence Whipple Observatory, Mt. Hopkins, AZ, USA. We acquired a total of 28 TRES observations of \thisstar{} with exposure times varying between 45 -- 540 seconds, achieving a median signal-to-noise per resolution element of 160. Our first observation was made on 2022-07-08; further observations were then delayed until October due to seasonal weather at Mt. Hopkins. We then regularly observed the star 17 times between 2022-10-04 and 2023-02-05 at an approximately weekly cadence. However, it became \bmaroon{apparent} that the expected large RV variability was not \bmaroon{present}, and the target was temporarily placed on hold. We subsequently acquired 10 further observations between 2023-10-04 and 2024-02-13, which confirmed the constant RV.

To determine the RVs, we perform spectral extraction and multi-order cross correlations following the process outlined in \citet{Buchhave2010}. We account for variations in the instrument zero-point through nightly observations of standard stars as in \citet{Zhou2023}. Our RV observations achieve a median precision of 11~\ms{}, and have a standard deviation of only 9~\ms{}.
We present the TRES RVs collected for this work in the Appendix.

In order to fully compare our RV results to the \textit{Gaia}~DR3 NSS astrometric solution, we require an accurate estimate for the stellar mass $M_*$. We therefore use the Stellar Parameter Classification tool \citep[SPC;][]{Buchhave2012} to estimate the observable stellar parameters from the TRES spectra as in \citet{Bieryla2024}. We estimate $T_{\text{eff}}=6000\pm50$~K, $[\text{M/H}]=-0.24\pm0.08$~dex, $\log g=4.26\pm0.10$~$\log$(c\ms{}), and a line broadening of $4.0\pm0.5$~k\ms{}. We then model the physical parameters of \thisstar{} using the MIST isochrones \citep{Dotter2016, Choi2016} with a model applied in our previous work \citep{Venner2024}. We use space-based photometry from Tycho-2, \textit{Gaia}, and 2MASS, and adopt the spectroscopic priors on effective temperature and metallicity from the SPC results. We estimate a stellar mass of $M_*=0.98^{+0.05}_{-0.04}~M_\odot$, radius $R_*=1.082\pm0.024~R_\odot$, luminosity $L_*=1.36\pm0.07~L_\odot$, and age $5.7^{+1.7}_{-1.9}$~Gyr. These properties suggest that \thisstar{} is similar to the Sun, mainly differing in being less metal-enriched and more evolved. We report our stellar parameters in Table~\ref{tab:parameters}.

With this updated value for the stellar mass, we may now recalculate the companion mass implied by the \textit{Gaia}~DR3 two-body solution using Equation~\ref{eq:astrometry}, resulting in $M_\text{p}=6.3\pm1.3~M_J$. Then, following Equation~\ref{eq:RV}, the expected reflex RV semi-amplitude is $K=170\pm35$~\ms{}.

We plot the zero-point normalised TRES radial velocities of \thisstar{} in Figure~\ref{fig:RV}. We compare it to the RV signal expected from the retracted \textit{Gaia}~DR3 two-body solution assuming the median orbital parameters. It \bmaroon{is} evident that our RV observations can unequivocally reject the existence of the candidate companion with high confidence, and could do so even prior to the official retraction of the orbital solution on 2024-05-27.

\section{Discussion and Conclusions} \label{sec:discussion}

The immediate result of this work is the non-detection of the (now-retracted) astrometric companion candidate of \thisstar{} originally reported in \textit{Gaia}~DR3. This is far from the first case of an astrometric exoplanet candidate that has gone undetected in follow-up RV observations; a notable pre-\textit{Gaia} example is VB~10 \citep{AngladaEscude2010, Bean2010}. However, in the vein of \citet{Rogers2024}, we seek here to leverage our non-detection to extract some lessons in the process of following up \textit{Gaia} astrometric exoplanet candidates for the forthcoming release of \textit{Gaia}~DR4.

\bmaroon{A major lesson drawn from past and present research is that} radial velocities provide a powerful complement to astrometry in the area of exoplanet detection. For solar-type FGK dwarfs, astrometry is mainly sensitive to long-period giant planets ($P\gtrsim1$~yr, $M_\text{p}\gtrsim1~M_J$) at the limits of current precision, which are normally well within the range of detectability of RVs. Furthermore, since astrometry achieves higher sensitivity for more proximate stars (Equation~\ref{eq:astrometry}), it performs best on nearby bright stars which are often the best RV targets.

Perhaps one of the main lessons to be taken from previous studies combining \textit{Hipparcos-Gaia} astrometry with radial velocities is that most bright solar-type stars have decades of RV survey observations fit to be combined with astrometry \citep[e.g.][]{Venner2021, Li2021, Xiao2023, An2025}. This is in significant part a result of the fact that most of the classical RV surveys assembled their target lists from bright solar-type stars in the \textit{Hipparcos} catalogue \citep[e.g.][]{Udry2000, Jones2002}, which is $>$90\% complete for stars brighter than $V\leq8$ \citep{Turon1992}. It is therefore unlikely that the \textit{Gaia}~DR4 astrometry will discover many \textit{new} planets orbiting solar-type stars brighter than $V\leq8$, simply due to the fact that most of these stars have already been observed in RV surveys. \bmaroon{Here \thisstar{} is an exception, being a Sun-like star omitted from the main northern hemisphere RV surveys. We speculate that this occurred due to its relatively early spectral type (F8). As a result, for the subsection of stars brighter than $V\lesssim8$, we argue that \textit{Gaia} may mainly discover new planets around earlier-type stars ($\lesssim$F8) typically unsuitable for precise RV observations due to their propensity for rapid rotation.}

Beyond $V>8$, the likelihood that a given solar-type star has existing RV observations declines due to a combination of the increasing likelihood of absence from \textit{Hipparcos} and idiosyncratic choices in target brightness limits among different RV surveys. \textit{Gaia} is therefore more likely to discover new planets around stars fainter than $V>8$, \bmaroon{which is} neatly demonstrated by Gaia-3~b, Gaia-4~b, Gaia-5~b\bmaroon{, and Gaia-6~B} \citep[all \bmaroon{$V>8$; e.g.}][]{Sozzetti2023, Stefansson2025, Pinamonti2026}, and \bmaroon{their} confirmation will entail new RV observations. Follow-up of the aforementioned targets have been undertaken by high-precision instruments \citep[i.e. $\approx$1~\ms{} precision; HIRES, HARPS-N, NEID, HPF;][]{Sozzetti2023, Stefansson2025}, which have proven adept for the confirmation of these \textit{Gaia}~DR3 planet candidates. However, access to observing time on high-precision spectrographs is restricted or highly competitive which may limit the accessibility of follow-up observations, especially considering the quantity of planet candidates expected in \textit{Gaia}~DR4 \citep{Perryman2014}.

We reason that there is a good analogy to be made between follow-up of \textit{Gaia}~DR4 planet candidates and the follow-up of exoplanets discovered by the TESS mission \citep{TESS}. TESS is performing an all-sky survey for transiting planets, predominantly those orbiting Sun-like stars between magnitudes $7\lesssim T\lesssim13$ (where $T$ reflects the red-optical TESS bandpass), which has detected thousands of planet candidates to date \citep{Guerrero2021}. The number, stellar properties, and magnitude distribution of TESS targets compare favourably to the properties of planet hosts expected to be discovered by \textit{Gaia} \citep{Perryman2014}. In the context of TESS science, intermediate-precision spectrographs ($\approx$10~\ms{} precision) have played a vital role in the follow-up and confirmation of TESS planet candidates through both outright confirmation of reflex RV signals as well as by excluding eclipsing binary false positive scenarios. In particular, we may highlight TRES, CHIRON \citep{Tokovinin2013}, MINERVA-Australis \citep{Addison2019}, and NRES \citep{Siverd2018}, along with the older FEROS \citep{Kaufer1999} and CORALIE \citep{Queloz2000}, as ``workhorse" spectrographs which have cumulatively contributed to the confirmation of hundreds of TESS exoplanets. These spectrographs typically achieve $10-100$~\ms{} RV precision for the relevant stars, down to $V\approx12$.

We argue that RV observations with comparable precision and breadth will play a key role in the follow-up of \textit{Gaia}~DR4 astrometric planet candidates. The hosts of \textit{Gaia} planets are expected to mainly lie between $6\lesssim G\lesssim16$, with a peak at $G\approx12$ \citep[where $G$ reflects the \textit{Gaia} bandpass;][figure~1a]{Perryman2014}. We infer from \citet{Perryman2014} that approximately $\approx$50\% of planets detected from the \textit{Gaia} astrometry will orbit stars between $8<G<12$, in the range amenable to observations by intermediate-precision spectrographs. While the conversion between the \textit{Gaia}-observed astrometric semi-major axis $a_0$ and the RV semi-amplitude $K$ depends on several parameters, (Equations~\ref{eq:astrometry}, \ref{eq:RV}), for a representative solar-mass star at 50~pc we suggest that typical values for $K$ may lie within an order of magnitude of $\approx$100~\ms{} for the planets expected to be detected by \textit{Gaia}\bmaroon{.} 
If such large values for $K$ are broadly accurate, a substantial fraction of \textit{Gaia} astrometric exoplanet discoveries could be directly confirmed using intermediate-precision spectrographs; for the remaining systems where $K$ is too small for detection, intermediate-precision spectrographs may still provide important information by providing vetting for false positives, especially binaries undetected in the \textit{Gaia} data.


Beyond the similarities in host astrophysical properties between TESS and \textit{Gaia} planet candidates, we further argue that TESS follow-up provides a reasonable model for the logistics of exoplanet follow-up for \textit{Gaia}~DR4. The TESS follow-up programme is primarily collaborative, which has helped greatly to maximise the scientific output of TESS. Given the even larger scale of exoplanet science expected from \textit{Gaia}, it can be argued that a similarly collaborative approach to \textit{Gaia} astrometric exoplanet follow-up would serve to enhance \textit{Gaia}~DR4 exoplanet science.

\vspace{10mm}

\section{Acknowledgements}

We acknowledge and pay respect to Australia’s Aboriginal and Torres Strait Islander peoples, who are the traditional custodians of the lands, waterways and skies all across Australia. We thank the anonymous referee for their comments that have helped to improve this work.
AVe would like to thank Jason Wang and Sarah Blunt for productive discussions on astrometric exoplanet detection that have helped to shape this work.
AVe and CXH are supported by ARC DECRA project DE200101840.

This work has made use of data from the European Space Agency (ESA) mission {\it Gaia} (\url{https://www.cosmos.esa.int/gaia}), processed by the {\it Gaia} Data Processing and Analysis Consortium (DPAC, \url{https://www.cosmos.esa.int/web/gaia/dpac/consortium}). Funding for the DPAC has been provided by national institutions, in particular the institutions participating in the {\it Gaia} Multilateral Agreement.


%

\vspace{5mm}
\facilities{\textit{Gaia},
            Tillinghast Reflector Echelle Spectrograph (TRES)}


\software{\texttt{nsstools} \citep{Halbwachs2023}}




\appendix


In Table~\ref{tab:RV} we present the multi-order, zero-point corrected TRES RVs collected for this work.

\startlongtable
\begin{deluxetable}{cccc}
\label{tab:RV}
\centering
\tablecaption{TRES radial velocity data for \thisstar{}.}
\tablehead{\colhead{BJD} & \colhead{RV (\ms{})} & \colhead{RV error (\ms{})}}
\startdata
2459768.970996 & -60.1 & 9.9  \\
2459856.822019 & -56.8 & 11.9 \\
2459862.845116 & -58.4 & 12.3 \\
2459872.911292 & -62.3 & 12.0 \\
2459880.849683 & -67.9 & 14.0 \\
2459888.778682 & -71.1 & 13.0 \\
2459896.807783 & -73.8 & 13.3 \\
2459905.797437 & -58.8 & 13.1 \\
2459913.755404 & -51.9 & 11.7 \\
2459922.725842 & -72.0 & 12.2 \\
2459928.834363 & -67.5 & 11.1 \\
2459932.793597 & -73.7 & 11.0 \\
2459941.617998 & -80.4 & 16.4 \\
2459949.773258 & -65.7 & 11.0 \\
2459958.680313 & -71.4 & 11.2 \\
2459969.665788 & -63.6 & 11.5 \\
2459975.590659 & -63.7 & 11.5 \\
2459980.592131 & -94.0 & 11.4 \\
2460221.964192 & -67.3 & 8.9  \\
2460222.909728 & -79.9 & 8.8  \\
2460235.903223 & -73.4 & 8.9  \\
2460249.743519 & -76.4 & 8.6  \\
2460263.840307 & -70.9 & 11.4 \\
2460282.702921 & -73.6 & 8.3  \\
2460295.725200 & -73.9 & 8.4  \\
2460310.719322 & -68.7 & 9.0  \\
2460326.699956 & -67.6 & 17.1 \\
2460353.594550 & -54.7 & 17.5 \\
\enddata
\end{deluxetable}



\bibliography{bib}
\bibliographystyle{mnras}



\end{document}